\documentstyle[prl,aps,preprint,tighten]{revtex}
\input{epsf}
\begin{document}
\draft
\preprint{CU-TP-902}
\title{Anomalous Chiral Symmetry Breaking above the QCD Phase
Transition}
\author{Shailesh Chandrasekharan\cite{Duke}, Dong Chen\cite{MIT},
Norman Christ, Weonjong Lee\cite{LANL}, Robert Mawhinney and
Pavlos Vranas} \address{Department of Physics, Columbia
University, \\
New York, New York 10027}
\date{July 8, 1998}
\maketitle
\begin{abstract}
We study the anomalous breaking of $U_A(1)$ symmetry just above
the QCD phase transition for zero and two flavors of quarks,
using a staggered fermion, lattice discretization.  The 
properties of the QCD phase transition are expected to depend on
the degree of $U_A(1)$ symmetry breaking in the transition
region.  For the physical case of two flavors, we carry out
extensive simulations on a $16^3\times 4$ lattice, measuring a
difference in susceptibilities which is sensitive to $U_A(1)$
symmetry and which avoids many of the staggered fermion
discretization difficulties.  The results suggest that anomalous
effects are at or below the 15\% level.
\end{abstract}
The breaking of classical $U_A(1)$ chiral symmetry by quantum
effects is a theoretical phenomenon of considerable physical
importance which has direct implications on the order of the
finite temperature, QCD phase transition.  If we consider only
the two light $u$ and $d$ quarks, anomalous symmetry breaking
reduces the flavor symmetry from $U(2)\times U(2)$ to $SU(2)
\times SU(2) \times U_B(1) \times Z_A(2)$.  It is only this
reduced symmetry that is consistent with second-order, critical 
behavior\cite{Pisarski-Wilczek}.  Although present lattice
simulations suggest that the QCD phase transition is indeed
second-order for two flavors, it is an important consistency
check to establish the required anomalous $U_A(1)$ symmetry
breaking directly.  In addition, the size of these symmetry
breaking effects will determine the width in temperature and
quark mass of the region showing universal critical behavior.

The plasma phase of QCD is a particularly good place to study the
axial anomaly.  For $T<T_c$, the dynamical breaking of chiral
symmetry obscures the effects of the axial anomaly.  For $T>T_c$,
chiral symmetry is restored and thermal Greens' functions are
explicitly symmetric under $SU(N_f)\times SU(N_f)$
transformations.  We can then look directly for anomalous
symmetry breaking by comparing Greens functions that are related
by the anomalous $U_A(1)$ symmetry.

Such breaking of global $U_A(1)$ symmetry, associated with a
zero-momentum Ward identity, is especially interesting, since at
zero momentum the anomalous term in the chiral Ward identity
becomes $N_f$ times the topological charge $\nu$, a quantity
which vanishes to all orders in conventional perturbation theory. 
The $\eta^\prime$ mass in QCD and the non-conservation of baryon
number in the standard model are other examples in which such
non-perturbative anomalous effects should occur\cite{tHooft}.

This sort of anomalous symmetry breaking can be understood from
two perspectives: i) As the physical remnant of an ultraviolet
ambiguity in the theory.  Here modifications made to regulate the
divergences present in the continuum gauge theory, necessarily
introduce explicit chiral symmetry breaking whose effects remain
visible, even on energy scales small compared to those of the
regulator.  ii) As resulting from an infrared singularity which
permits chiral asymmetry to survive the chiral limit of vanishing
fermion mass.  While chiral symmetry breaking effects might
naively be expected to be proportional to the explicit quark
mass, $m$, the presence of infrared singular, $1/m$ behavior can
allow such chirally asymmetric effects to remain even in the
$m\rightarrow 0$ limit.  In a semiclassical calculation, such
$1/m$ behavior appears for gauge backgrounds with fermion zero
modes.  These two views are related by the Atiyah-Singer index
theorem\cite{Brown}.

Most lattice calculations\cite{Yukuwa-Kilcup} explore the first
approach, studying explicit chiral symmetry breaking inherent in
the lattice regulation which should reduce to anomalous effects
as the continuum limit is taken\cite{Karsten-Smit}.  In this
paper we explore the second approach, searching for anomalous
asymmetries that arise from infrared singularities in the limit
of small quark mass.  For lattice calculations these two
approaches are not equivalent since the Atiyah-Singer theorem
applies only in the limit of an infinite number of degrees of
freedom.  

It is important to pursue both methods.  The first approach may
over-estimate anomalous effects, confusing them with simple
lattice artifacts which vanish in the continuum limit.  The
second approach may miss anomalous effects since infrared
singularities are often softened by lattice effects, {\it e.g.}
the zero-mode-shift of Smit and Vink\cite{Smit-Vink}, and become
apparent only as $a\rightarrow 0$.

The question of anomalous symmetry breaking above $T_c$ has now
been studied by a number of groups.  For a general review see the
article of Laermann\cite{Laermann}.  Preliminary versions of our
results can be found in Ref.~\cite{Chandra}, while an alternative
calculation, also taking approach ii), can be found in Bernard,
{\it et al.}\cite{Bernard}.  Finally, Kogut, {\it et
al.}\cite{Kogut} use a combination of methods examining signals
for anomalous symmetry breaking of both types i) and ii) above.

In this paper, we study both zero- and two-flavor QCD, just above
$T_c$.  For $N_f=0$, anomalous effects are expected in the chiral
condensate, $\langle\bar q q \rangle$, while for $N_f=2$ we must
examine a more infra-red singular, $U_A(1)$-noninvariant
quantity, here a difference of iso-vector susceptibilities which
we refer to as $\omega = \chi_P -\chi_S$ where
\begin{equation}
\chi_P={1 \over 2\Omega}\int d^4x\, d^4y \langle \bar\chi\tau^j i
\gamma^5\chi(x) 
     \bar\chi\tau^j i \gamma^5\chi(y) \rangle
\label{eq:prop}
\end{equation}
for space-time volume $\Omega$ and flavor generator $\tau^j$. 
The scalar susceptibility $\chi_S$ is defined similarly, by
omitting the internal $i\gamma^5$ factors.

We adopt the staggered fermion, lattice discretization.  This is
the approach used most successfully to date in finite
temperature, lattice QCD studies.  While the quantity
$\langle\bar q q\rangle$ can be calculated directly using this
formalism, the more physically interesting $\omega = \chi_P -
\chi_S$ can not.  Direct definitions of $\chi_P$ and $\chi_S$
using staggered fermions will necessarily introduce ambiguities,
with potentially large lattice artifacts obscuring the anomalous
effects of interest.

Here, we take an indirect approach, expressing $\omega$, in the
continuum, as a spectral integral whose singular behavior as
$m\rightarrow 0$ gives rise to anomalous symmetry breaking.  We
then demonstrate that this spectral integral can be directly
evaluated using staggered lattice fermions and use this result to
provide a lattice calculation of $\omega$.

Consider the spectral representations:
\begin{mathletters}
\label{eq:spectral}
\begin{eqnarray}
\label{eq:banks-casher}
\langle\bar q q\rangle &=& -2m_\zeta \int_0^\infty d\lambda 
   {\rho(\lambda, g^2, m)\over\lambda^2 + m_\zeta^2 } 
   \Bigr|_{m_\zeta=m}\\
\omega                 &=& 4 m^2 \int_0^\infty d\lambda
   {\rho(\lambda, g^2, m)\over(\lambda^2+m^2)^2}.
\label{eq:omega_spect}
\end{eqnarray}
\end{mathletters}
Here $\rho(\lambda, g^2, m)$ is the average density of Dirac
eigenvalues $\lambda$.  The first formula is due to Banks and
Casher\cite{Banks-Casher} and the second is derived in a similar
fashion.  In Eq.~\ref{eq:banks-casher} we distinguish the fermion
mass that appears in the fermion line attached to $q$ and $\bar
q$, $m_\zeta$, from that entering through the fermion
determinant, $m$.  The factors of $m$ or $m_\zeta$ in the
numerators of Eq.~\ref{eq:spectral} reflect the chiral symmetry
breaking character of $\langle\bar q q \rangle$ and $\omega$. 
However, an anomalous, small-mass limit can result if the
integral over $\lambda$ is sufficiently singular for small
$\lambda$.

Now let us investigate what might be expected for these
quantities in continuum QCD.  For $T>T_c$, the small mass limit
of $\langle\bar q q \rangle$ and $\omega$ in the continuum theory
can be analyzed for both the case of very small volume and in the
limit of infinite volume.  For finite volume, the Dirac spectrum
will be discrete for each gauge configuration in the path
integral.  The only non-zero contributions to either $\langle\bar
q q \rangle$ or $\omega$ as $m \rightarrow 0$ will come from
gauge configurations with at least one exact Dirac zero mode.  In
very small volumes, these zero modes can be predicted
semiclassically and give the anomalous, small-mass behaviors:
$\langle\bar q q \rangle \sim 1/m$, for $N_f=0$ and $\omega \sim
$ constant, for $N_f=2$.

The case where $V\rightarrow\infty$ first is more interesting and
can be analyzed using the methods of Leutwyler and
Smilga\cite{Smilga}.  Above $T_c$, there are no massless modes so
the free energy should be proportional to the volume and analytic
in the fermion mass:
\begin{equation}
Z \approx \exp \Omega \bigl\{F_0 + F_2 {\rm tr} M^\dagger M +
G\,{\rm re}\{e^{i\theta} {\rm det} M\}\bigr\},
\label{eq:free-energy}
\end{equation}
where $M$ is the complex fermion mass matrix and $\theta$ the
usual theta parameter.  Defining the topological susceptibility
as $\chi_{\rm top} = -\partial^2/\partial \theta^2 \ln{Z}$, one
easily derives $\chi_{\rm top} = \Omega G\, m^{N_f}$ for
$\theta=0$ and $M=m$I, a multiple of the identity.

We can similarly obtain expressions for $\langle\bar q q\rangle$
and $\omega$:
\begin{mathletters}
\label{eq:smilga}
\begin{eqnarray}
\langle\bar q q\rangle&=& -{1\over N_f\,\Omega}     
     {\partial\over\partial m} \ln{Z(M)}   
                         = -2 F_2 m - G m^{N_f-1} 
\label{eq:replica} \\
\omega&=& {1 \over \Omega} 
     \{{\partial^2 \over \partial m_{\rm r}^{j\;2}} -
       {\partial^2 \over \partial m_{\rm i}^{j\;2}}\}\ln{Z(M)}
               = 2 G m^{N_f-2}
\label{eq:complex}
\end{eqnarray}
\end{mathletters}
where in Eq.~\ref{eq:replica} we have divided by $N_f$ to define
$\langle\bar q q\rangle$ as coming from a single fermion specie
while in Eq.~\ref{eq:complex} we have used a complex $M=m{\rm
I}+(m_{\rm r}^j+i\,m_{\rm i}^j)\tau^j$.

If we make the possibly reasonable assumption that the quenched
value of $\langle\bar q q\rangle$ can be obtained as the
$N_f\rightarrow 0$ limit of Eq.~\ref{eq:replica}, then we can
combine Eq.~\ref{eq:smilga} with the formula for $\chi_{\rm top}$
to obtain:
\begin{equation}
\begin{array}{llll}
N_f=0: \quad &\langle\bar q q\rangle &= -{1 \over m}  {\chi_{\rm
top}\over \Omega} & \sim {1\over m}\\

N_f=2: \quad &\omega                 &= {2 \over m^2}{\chi_{\rm
top}\over \Omega} & \sim {\rm const.}
\end{array}
\label{eq:infinitevol}
\end{equation}
The last relation is of particular interest, implying that above
$T_c$ the quantity $\omega$ provides an alternative measure of
the topological susceptibility.   As is shown below, $\omega$ can
be easily determined using lattice methods, without the normal
difficulties of defining topological winding on a discrete
lattice.

We will now compare these continuum expectations with lattice
calculations.  Because of the remnant chiral symmetry of
staggered fermions, Eq.~\ref{eq:banks-casher} is also valid on
the lattice, allowing us to relate $\langle\bar\chi\chi\rangle$
and $\rho$, where $\chi$ is the single component, staggered
fermion field.  Viewing $\langle\bar q q\rangle$ as a function of
$m$ and $m_\zeta$, we can express $\omega$ as a function of
$\langle\bar q q\rangle$ and then use this continuum result to
define $\omega$ on the lattice:
\begin{equation}
\label{eq:lattice}
\omega = -{1\over m}\langle\bar\chi\chi\rangle
          +{\partial \over \partial m_\zeta} 
              \langle\bar\chi\chi\rangle|_{m_\zeta=m}
\end{equation}
where these two terms correspond precisely to the terms in the
difference $\omega=\chi_P-\chi_S$.  In the remainder of this
paper we quote values of $\omega$, $\chi_P$ and $\chi_S$
normalized according to Eq. \ref{eq:lattice} where
$\langle\bar\chi\chi\rangle$ is normalized to behave as $1/m$ in
the large mass limit.

\begin{figure}[p]
\centering
\epsfxsize=5.5in
\centering
\leavevmode
\epsfbox{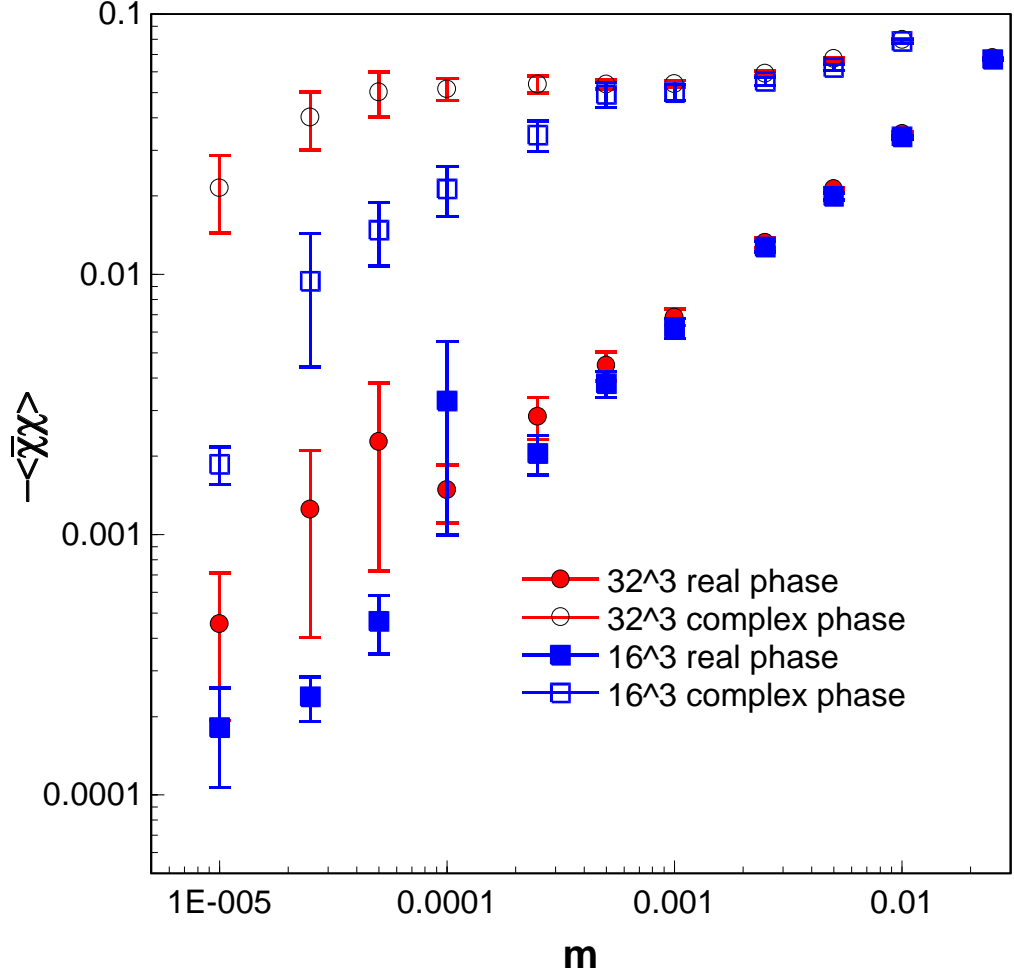}
\caption{The chiral condensate $\langle \bar\chi\chi \rangle$
plotted as a function of quark mass for a pure gauge calculation
on $16^3 \times 4$ and $32^3 \times 4$ lattices.  The real phase
(closed points) is the most physical (${\rm det}(D-m)$ is largest
for this phase).  No evidence is seen for the expected anomalous
behavior,  $\langle \bar\chi\chi \rangle \sim m^{-1}$ as
$m\rightarrow 0$.}
\label{fig:b5_71_4}
\end{figure}

First consider $N_f=0$.  In Fig.~\ref{fig:b5_71_4} we show
$\langle\bar\chi\chi\rangle$ for two distinct phases
distinguished by the complex phase of the Wilson line, $\langle W
\rangle$, computed at $\beta=5.71$, just above $\beta_c=5.6925$. 
(Recall the Wilson line, $W$, is the volume average of the trace
of the ordered product of link variables along a line in the time
direction.)  For the case where $\langle W \rangle$ is real, we
see the power law $\sim m^{0.76}$ for both $16^3$ and $32^3$
volumes suggesting this power law description holds in the
infinite volume limit.  We see no sign of the anomalous $1/m$
behavior in $\langle\bar\chi\chi\rangle$ expected in the
continuum.

For the case of complex $\langle W \rangle$, we see an unexpected
spontaneous breaking of chiral symmetry above $T_c$, with
$\langle\bar\chi\chi\rangle$ approaching a constant as $m$
decreases.  The eventual decrease in $\langle\bar\chi\chi\rangle$
for very small $m\le m_{\rm min}$ is the normal finite-volume
behavior expected with spontaneous symmetry breaking, with
$m_{\rm min}\langle\bar\chi\chi\rangle V/T \approx 1$ for both
volumes.

\begin{figure}
\epsfxsize=6.0in
\centering
\leavevmode
\epsfbox{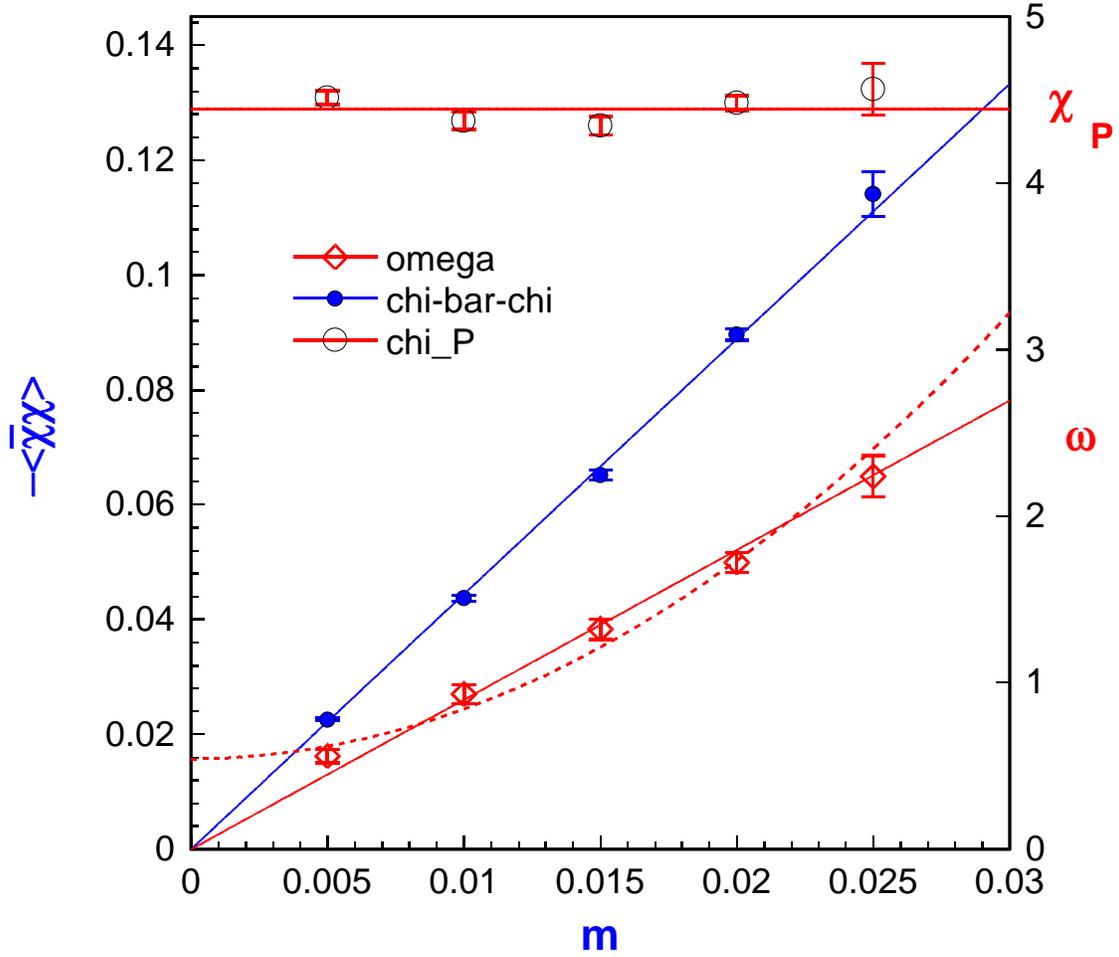}
\caption{ The quantity $\omega$, which directly measures
anomalous symmetry breaking, plotted versus fermion mass, $ma$. 
Also shown are the chiral condensate $\langle\bar\chi\chi\rangle$
and the pseudoscalar susceptibility $\chi_{\rm P}$.  We studied a
$16^3\times 4$ lattice at $\beta=5.3$, just above $\beta_c$.}
\label{fig:omega}
\end{figure}

Next we examine $\omega$ and the more physical case of two
flavors, at $\beta=5.3$, just above $\beta_c$ (recall $\beta_c
\approx 5.265$ for $N_t=4$ and $ma$=0.01), on a $16^3\times 4$
lattice for five different values of the dynamical quark mass. 
The results are summarized in Table~\ref{tab:data} and plotted in
Fig.\ref{fig:omega}.  This figure shows the chiral condensate,
$\langle\bar\chi\chi\rangle$ approaching zero linearly as is
expected for $\beta>\beta_c$.  Likewise, $\chi_P$ shows the
expected regular, constant behavior as $m\rightarrow 0$. 
However, rather than showing the anomalous behavior, $\omega \sim
\omega_0 + \omega_2 m^2$, expected from Eq.~\ref{eq:infinitevol},
Fig.\ref{fig:omega} suggests a nearly linear $\omega$ as $m$ goes
to zero.

Four fitted curves are also shown in Fig.~\ref{fig:omega}.  The
two linear fits to $\langle\bar\chi\chi\rangle$ and $\omega$ have
a $\chi^2/dof$ of 2.2 and 2.7 respectively.  Both of these fits
are constrained to vanish at $m=0$.  If that constraint is
dropped for the $\omega$ fit, the intercept moves upward slightly
to 0.15(5) and the $\chi^2/dof$ falls to 0.34.  A fit to the
expected form $\omega_0 + \omega_2 m^2$ is worse, with a
$\chi^2/dof$ of 3.4.

While these results are consistent with those reported by Bernard
{\it et al.}\cite{Bernard}, our conclusions are different.  That
calculation examines a smaller lattice spacing than considered
here but with larger statistical errors.  Their analysis adopts
the quadratic small-mass dependence for $\omega$.  However, such
quadratic behavior is only guaranteed theoretically in the
unphysical limit where $m$ vanishes at fixed lattice spacing and
needs to be established numerically for the case of interest.

From Fig.~\ref{fig:omega} one observes that for quark masses in
the range $0.01 \le m \le 0.025$, our results are consistent with
an unusual but non-anomalous, linear behavior $\omega \sim m$. 
At our smallest mass, 0.005, $\omega$ is significantly higher
than such a linear extrapolation, suggesting that anomalous
effects may be emerging.  However, such effects are clearly quite
small and occur for quark masses that are below those used in
present studies of QCD thermodynamics, suggesting little
connection between this anomalous behavior and the observed
second-order QCD phase transition.  

In order to describe the physical size of a possible non-zero
value of $\omega|_{m=0}$, we must address the potential cut-off
dependence of the quantities being discussed.  While a thorough
analysis of this question lies beyond the scope of the present
paper\cite{long-paper}, there are two issues that are important
to recognize.  First, the $m=0$ intercept of $\omega$ requires
the same $\ln(a)$-dependent, multiplicative renormalization as
the inverse square of the quark mass $m$, as is suggested by
Eq.~\ref{eq:infinitevol}.  We will ignore such a factor for our
present rough estimate, since this factor should be of order 1
for current lattice spacings.  

Since $\omega$ is the difference of $\chi_P$ and $\chi_S$, it is
natural to compare $\omega$ to either of these quantities. 
However, both quantities contain a $1/a^2$ piece when evaluated
in physical units.  Thus, we choose to compare $\omega$ to a
quantity we will call $\tilde\omega$, obtained as the difference
between $\chi_P$ evaluated at $\beta=5.3$ and $\chi_S$ evaluated
at $\beta=5.245$, just below the transition:
\begin{equation}
\tilde\omega={\langle\bar\chi\chi\rangle \over m}
                                              \big|_{\beta=5.3}
            -{\partial \langle\bar\chi\chi\rangle 
                   \over \partial m_\zeta}\big|_{\beta=5.245}.
\end{equation}
Such a subtraction removes the unwanted, quadratically divergent
$1/a^2$ term at tree level and leaves an expression finite up to
an $\cal O$(1), $\ln(a)$-dependent multiplicative factor and a
much smaller $1/a^2$ term suppressed by the factor $(5.3-5.245)$. 
We find $\omega/\tilde\omega \sim 15\%$.

Within the expected critical region, the light modes ($\vec\pi$,
$\sigma$) should be much less massive than the non-universal
degrees of freedom suggesting $\omega/\tilde\omega \sim 1$, not
the $\sim 0.15$ observed here.  Thus, our results suggest that
$O(4)$ critical behavior should not be seen in $N_t=4$
thermodynamics at least for $|\beta-\beta_c| \approx 0.03$.

In conclusion, we have numerically studied anomalous symmetry
breaking by examining quantities whose anomalous behavior comes
directly from infrared effects.  Given the relatively coarse
lattice spacing in our simulations $a \approx $1 Fermi, our
failure to find such effects above the 15\% level is far from
conclusive evidence that such effects are suppressed in
Nature\cite{Kaehler}.  However, this represents a first step in a
systematic lattice calculation of such phenomena and must be
followed by more demanding calculations on finer lattices and
calculations using fermion formulations with improved chiral
properties.

We thank Edward Shuryak and Andre Smilga for helpful discussions
and Yubing Luo for assistance.

\begin{table}
\caption{Our $\beta=5.3$, $16^3\times 4$ results for two flavors
of dynamical quark with mass $m$.  Run length is the number of
time units in the hybrid, `R'-algorithm evolution after 200 time
units were discarded.  These quantities are normalized in a
manner consistent with Eq.~\protect\ref{eq:lattice} with
$\bar\chi\chi$ defined so that it behaves as $1/ma$ for large
$ma$.}
\label{tab:data}
\begin{tabular}{lclll}
\multicolumn{1}{c}{$ma$}
      &\multicolumn{1}{c}{run length}
               &\multicolumn{1}{c}{$\langle\bar\chi\chi\rangle$}
                              &\multicolumn{1}{c}{$\chi_S$}
                                  &\multicolumn{1}{c}{$\omega$}\\
0.005 & 4464   & 0.02256(21)  & 3.945(43) & 0.559(37) \\
0.01  & 2550   & 0.04374(50)  & 3.369(68) & 0.932(56) \\
0.015 & 2600   & 0.06517(82)  & 3.010(52) & 1.322(62) \\
0.02  & 2992   & 0.0896(10)   & 2.697(34) & 1.722(62)  \\
0.025 & 3072   & 0.1141(38)   & 2.38(11)  & 2.24(12)  \\
\end{tabular}
\end{table}

\end{document}